\documentclass[twocolumn,nofootinbib,amsmath,amssymb,a4paper]{revtex4}

\usepackage{graphicx}
\usepackage{dcolumn}
\usepackage{bm}

\newcommand\babar{\mbox{\slshape B\kern-0.1em{\small A}\kern-0.1em
    B\kern-0.1em{\small A\kern-0.2em R}}}
\newcommand{\brhoks}{$B \rightarrow \rho {K^*}$}
\newcommand{\brhozkp}{$B^+ \rightarrow \rho^0 {K^*}^+$}
\newcommand{\brhomkp}{$B^0 \rightarrow \rho^- {K^*}^+$}
\newcommand{\brhopkz}{$B^+ \rightarrow \rho^+ {K^*}^0$}
\newcommand{\brhozkz}{$B^0 \rightarrow \rho^0 {K^*}^0$}
\newcommand{\bfzks}{$B \rightarrow f_0(980) {K^*}$}
\newcommand{\bfzkp}{$B^+ \rightarrow f_0(980) {K^*}^+$}
\newcommand{\bfzkz}{$B^0 \rightarrow f_0(980) {K^*}^0$}
\newcommand{\bomks}{$B \rightarrow \omega {K^*}$}
\newcommand{\bomkz}{$B^0 \rightarrow \omega {K^*}^0$}
\newcommand{\bomkp}{$B^+ \rightarrow \omega {K^*}^+$}
\newcommand{\bomrh}{$B \rightarrow \omega \rho$}
\newcommand{\bomrz}{$B^0 \rightarrow \omega \rho^0$}
\newcommand{\bomrp}{$B^+ \rightarrow \omega \rho^+$}
\newcommand{\bomom}{$B^0 \rightarrow \omega \omega$}
\newcommand{\bomph}{$B^0 \rightarrow \omega \phi$}
\newcommand{\bomfz}{$B^0 \rightarrow \omega f_0(980)$}
\newcommand{\rkp}{$\rho K \pi$}
\newcommand{\BB}{$B\bar{B}$}
\newcommand{\mpp}{${\rm m}_{\pi \pi}$}
\newcommand{\mkp}{${\rm m}_{K \pi}$}
\newcommand{\Mpp}{${\rm M}(\pi^+ \pi^0)$}
\newcommand{\Mkp}{${\rm M}(K^+ \pi^-)$}
\newcommand{\mbc}{${\rm M}_{\rm bc}$}
\newcommand{\mes}{${\rm m}_{\rm ES}$}
\newcommand{\DE}{$\Delta \rm{E}$}
\newcommand{\fL}{$f_L$}
\newcommand{\Acp}{${\cal A}_{CP}$}

\begin{document}

\title{\brhoks\ decays and other rare vector-vector modes
\footnote{Presented at the $4^{th}$ International Workshop 
on the CKM Unitariry Triangle, Nagoya, Japan, December 12-16, 2006.
Preprint DAPNIA-06-601.}}

\author{G. Vasseur}
 \email{georges.vasseur@cea.fr}
\affiliation{DSM/DAPNIA/SPP, CEA/Saclay,
F-91191 Gif-sur-Yvette, France}

\begin{abstract}
The recent analyses of the following rare vector-vector decays of the $B$ meson
are presented:
$\rho K^*$, $\omega K^*$, $\omega \rho$, $\omega \omega$, and $\omega \phi$
charmless final states.
The latest results indicate that the fraction of longitudinal polarization 
is about 0.5 in penguin-dominated modes 
and close to 1 for tree-dominated modes.
\end{abstract}

\maketitle

\section{Motivation}
The search for rare charmless hadronic decays of the $B$ meson 
to vector-vector final states has become a quite active field 
in the experiments at the $B$ factories, Belle at KEK and \babar\ at SLAC. 
As a lot of these decays have not yet been seen,
the first goal of these studies is to observe such modes
and measure their branching fraction.
The measurements can then be compared to theoretical predictions.

The direct CP-violation asymmetry in these modes can also be measured.
It is defined as 
${\cal A}_{\rm CP} = (\Gamma^- - \Gamma^+)/(\Gamma^- + \Gamma^+)$, 
where the superscript on the total width $\Gamma$ indicates the sign of 
the $b$-quark charge in the $B$ meson.
Some modes can be used for further CP studies.
In fact, the result on \brhopkz\ has already been used 
to constrain the effect of the penguin amplitude 
on the measurement of the angle $\alpha$ of the unitarity triangle
from $B^0 \to \rho^+ \rho^-$ using SU(3) flavor symmetry~\cite{bib:beneke}.

\begin{figure}
\includegraphics[width=0.4\textwidth]{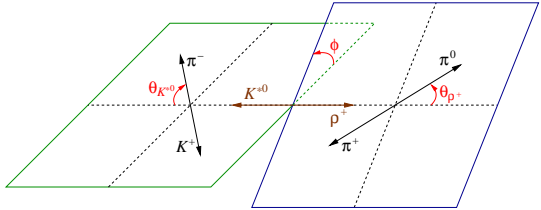}
\caption{\label{fig:angles}Definition of the helicity angles 
in the case of the vector-vector \brhopkz\ decay.}
\end{figure}

A hot topic is the measurement of the fraction of longitudinal polarization. 
The helicity angles $\theta_1$ and $\theta_2$ of the two vector mesons 
are defined
as the angles between the vector meson direction in the $B$ meson rest frame
and the direction of one of its decay product in the vector meson rest frame,
as illustrated on one example in Fig.~\ref{fig:angles}.
Integrating over the $\phi$ angle 
between the decay planes of the two vector mesons,
the fraction of longitudinal polarization \fL\ can be extracted 
from the angular dependence of the decay rate, which is proportional to
\begin{equation}
\frac{1}{4}
(1-f_L)\sin^2\theta_{1}\sin^2\theta_{2} + 
   f_L \cos^2\theta_{1}\cos^2\theta_{2} . \nonumber
\label{eq:helicity}
\end{equation}

A value of \fL\ close to unity of order $(1 - O(\frac{{m_V}^2}{{m_B}^2}))$ 
is expected for light vector mesons from helicity conservation.
This is expected to be true for both tree and penguin diagrams.
However the experimental situation is more complex.
If \fL\ has indeed been measured close to 1 
in the tree-dominated $B \to \rho \rho$ modes~\cite{bib:somov},
it is surprisingly close to 0.5 
in the penguin-dominated $B \to \phi K^*$ modes~\cite{bib:chen}. 
This effect is not yet understood.
There are several possible explanations, either within the Standard Model,
such as rescattering in the final state, 
contribution from annihilation or electroweak penguin diagrams,
and transverse gluon~\cite{bib:hou}, 
or in new physics outside the Standard Model.
To have a better picture, it is important to measure other vector-vector modes,
both tree-dominated, like \bomrh\ and \bomom, 
and penguin-dominated like \brhoks\ and \bomks.

The recent studies of the \brhoks\ modes are reviewed in section~\ref{sec:rk},
the ones involving an $\omega$ meson in section~\ref{sec:om}.
Charge-conjugate modes are implied throughout.

\section{\brhoks\ modes}
\label{sec:rk}

\subsection{Introduction}

\begin{figure}[h]
\includegraphics[width=0.4\textwidth]{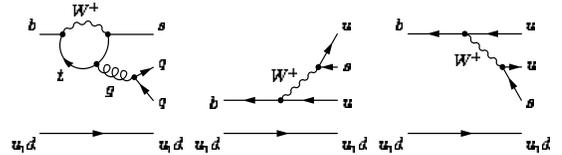}
\caption{\label{fig:diag}Feynmann diagrams for the \brhoks\ decay:
gluonic penguin, external tree and internal tree diagrams.}
\end{figure}

The \brhoks\ charmless decays proceed through dominant gluonic penguin loops 
and doubly Cabibbo-suppressed tree processes, as shown in Fig.~\ref{fig:diag}.
The external tree diagram is only possible with a $K^{*+}$,
and the color-suppressed internal tree diagram with a $\rho^0$.
Hence \brhopkz\ is pure penguin.

According to isospin symmetry, the two modes with a charged $\rho$ 
are expected to have a branching fraction twice as large 
as the two modes with a neutral $\rho$.

\subsection{Results from Belle}

\begin{figure}[h]
\includegraphics[width=0.2\textwidth]{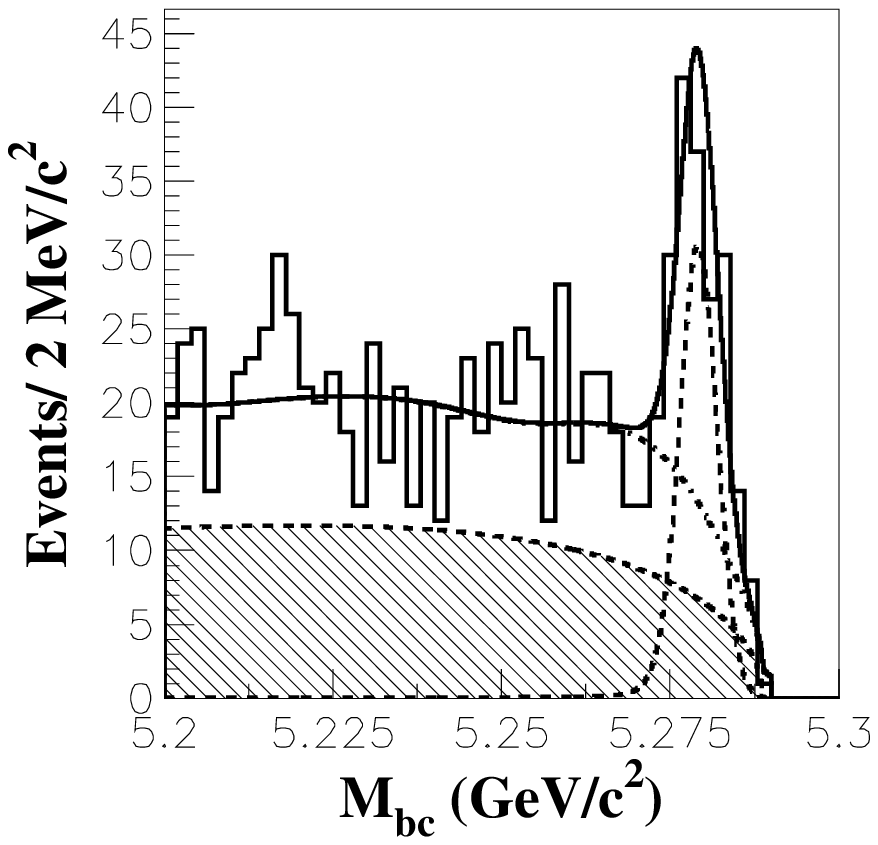}
\includegraphics[width=0.2\textwidth]{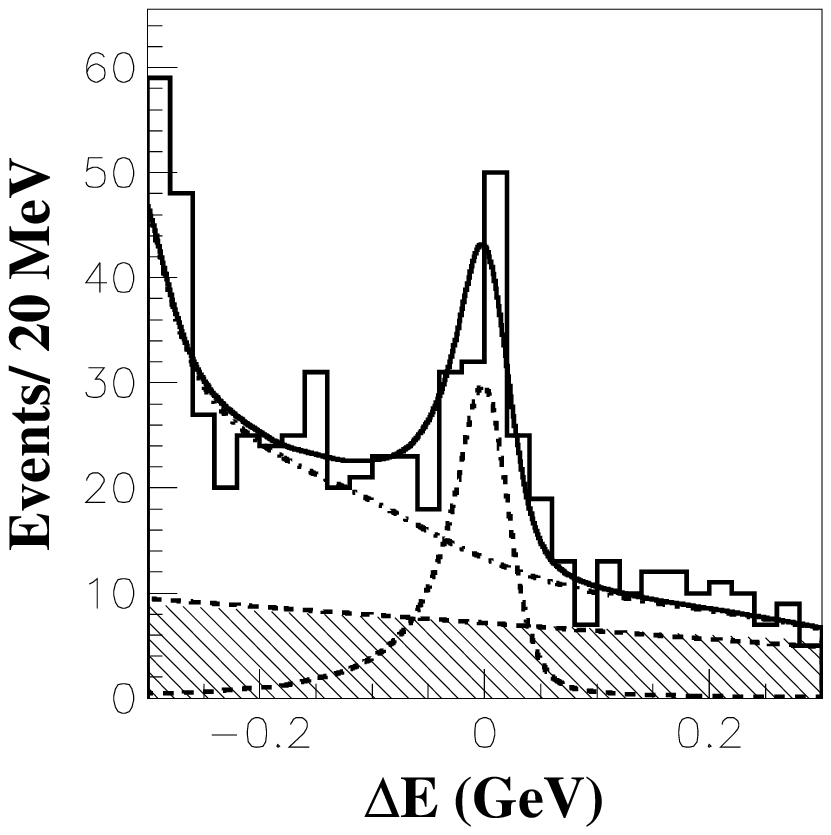}
\caption{\label{fig:belle_mbcde}Projections of \mbc\ for events in the 
\DE\ signal region (left) and of \DE\ in the \mbc\ signal region (right).
The solid curves show the results of the fit.
The dashed curve is the signal contribution.
The hatched histograms represent the continuum background. 
The sum of the $b\to c$ and continuum background component is shown 
as dot-dashed lines.}
\end{figure}

\begin{figure}[h]
\includegraphics[width=0.2\textwidth]{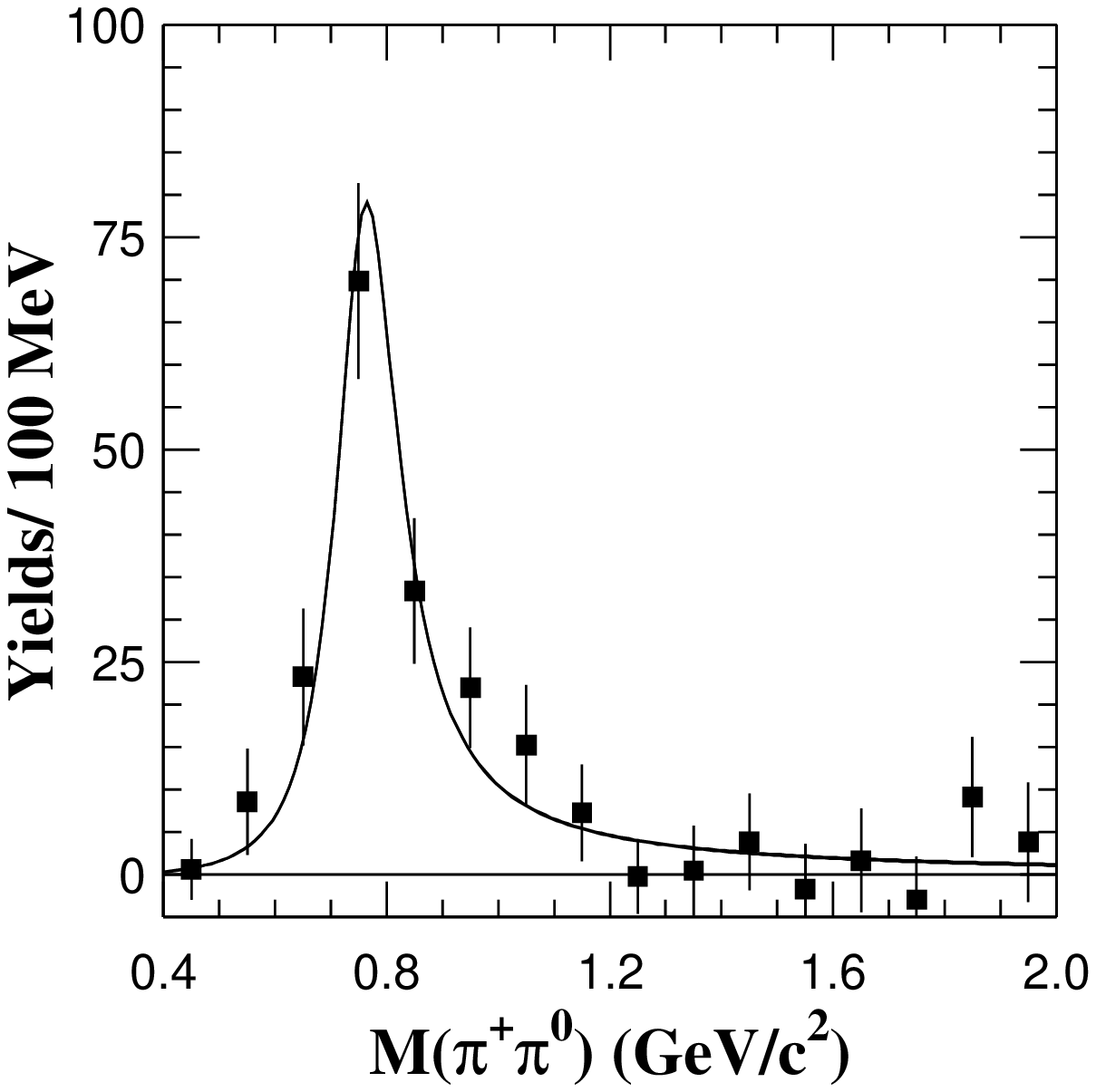}
\includegraphics[width=0.2\textwidth]{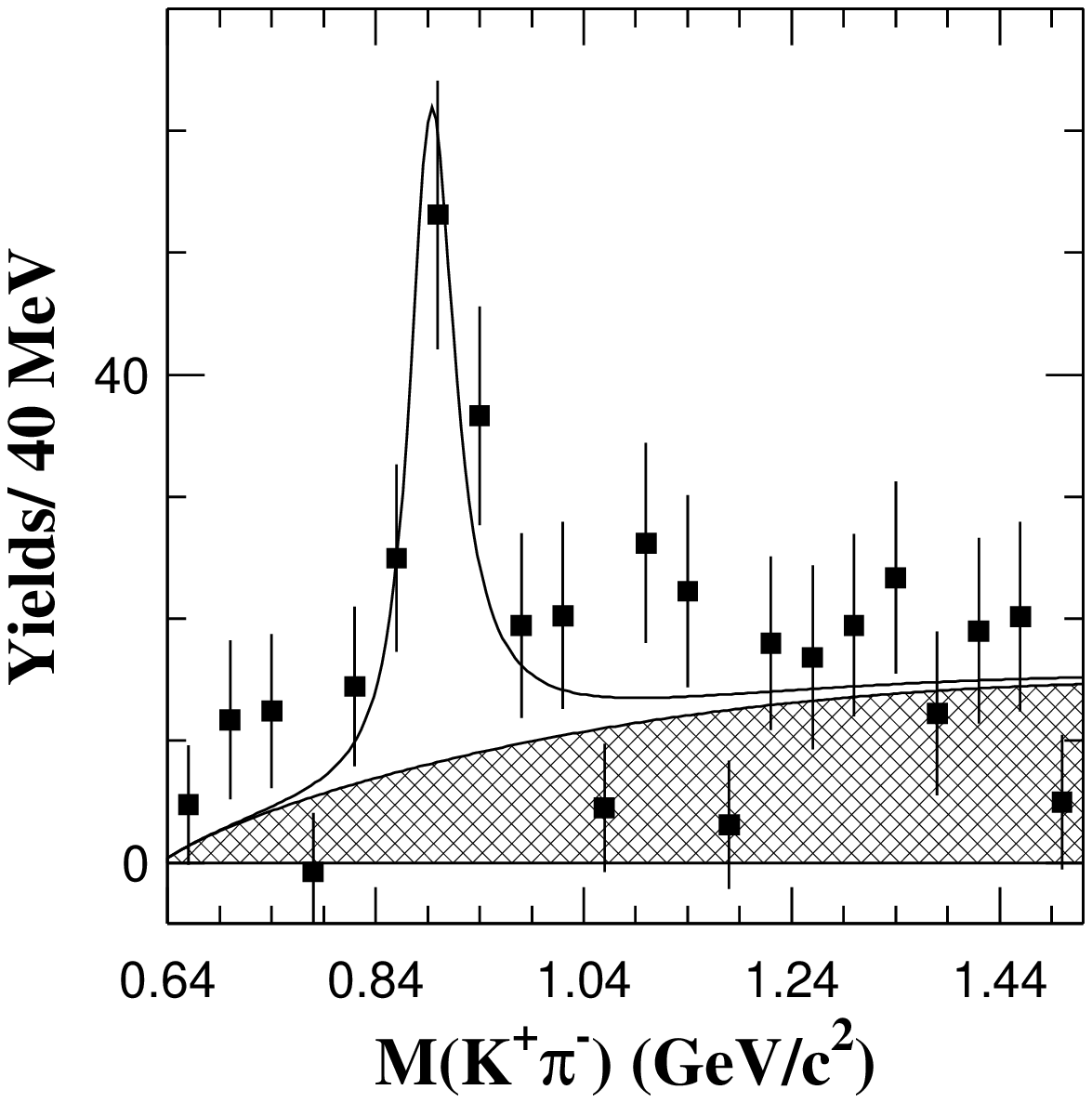}
\caption{\label{fig:belle_mppmkp}Signal yields obtained 
from the \mbc-\DE\ distribution in bins of \Mpp\ (left) 
for events in the $K^{*0}$ region and in bins of \Mkp\ (right)
for events in the $\rho^+$ region.
The points with error bars show the data.
Solid curves show the results of the fit.
Hatched histograms are for the nonresonant component.}
\end{figure}

Belle was the first experiment in 2005 to publish a result 
on the observation of the \brhopkz\ mode~\cite{bib:belle},
on a sample of 275 millions of \BB\ pairs.
A signal of $B^+ \to \pi^+ \pi^0 K^+ \pi^-$ is extracted
from the $e^+ e^- \to q \bar{q}$ continuum and \BB\ backgrounds  
in an extended unbinned maximum-likelihood fit using
the $B$ meson beam-constrained mass \mbc\ and energy difference \DE,
as shown in Fig~\ref{fig:belle_mbcde}.

The \brhopkz\ signal is extracted by fits to \mbc\ and \DE\ 
in bins of the vector meson masses \Mpp\ and \Mkp,
as shown in Fig~\ref{fig:belle_mppmkp}.
This is necessary because there is a large nonresonant \rkp\ background,
which gives a continuum in the distribution of \Mkp. 
Nethertheless there is a clear \brhopkz\ signal of $85\pm16$ events 
with a significance of 5.2 $\sigma$.

As for \fL, it is obtained by fitting simultaneously
the signal yields obtained from \mbc-\DE\ fits in bins of
the two helicity angles,
assuming an S-wave $K \pi$ system in the \rkp\ background.
The results for the branching fraction and \fL\ in \brhopkz\ are:
\begin{eqnarray}
    {\cal B} &=& (8.9\pm1.7\pm1.2) \, 10^{-6},  \nonumber \\
    f_L &=& 0.43\pm0.11^{+0.05}_{-0.02}. \nonumber 
\end{eqnarray}
The value found for \fL\ is similar to the one found in $\phi K^*$
and its error is about twice as large as in $\phi K^*$.

\subsection{Results from \babar}

\begin{figure}[h]
\includegraphics[width=0.4\textwidth]{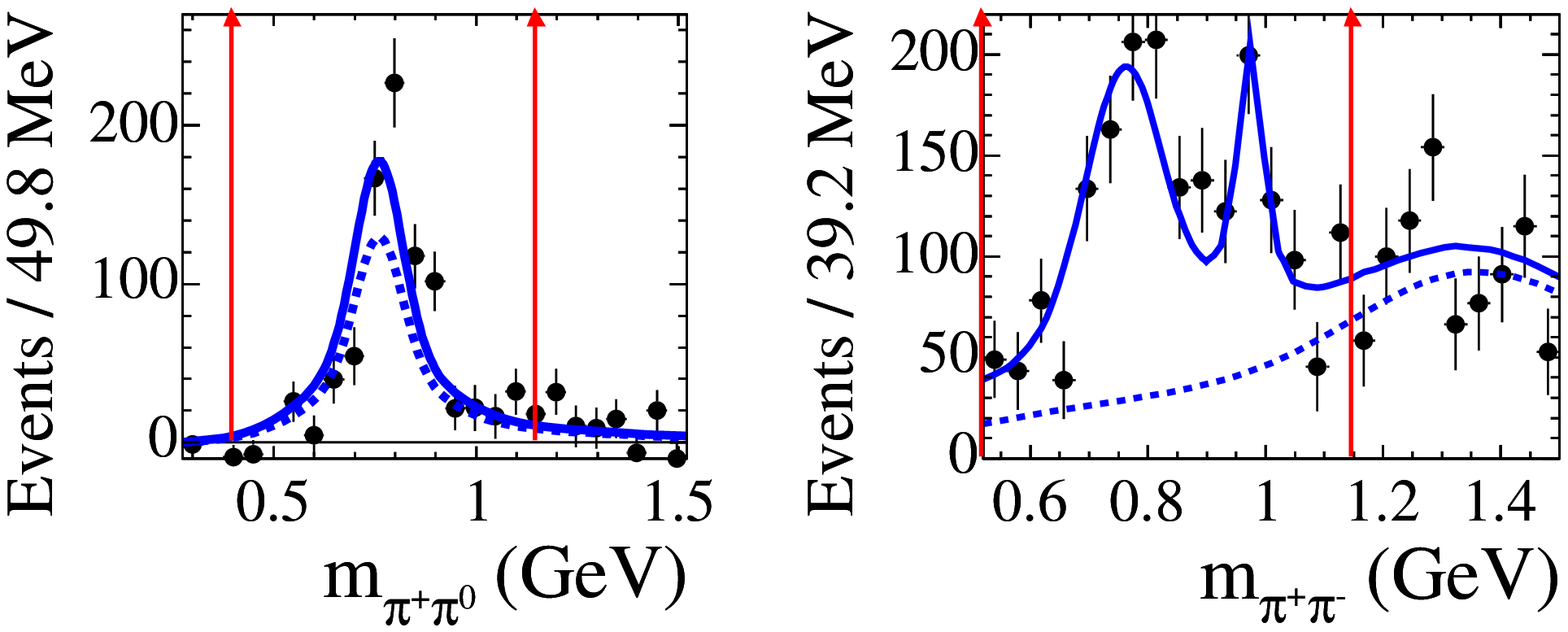}
\includegraphics[width=0.4\textwidth]{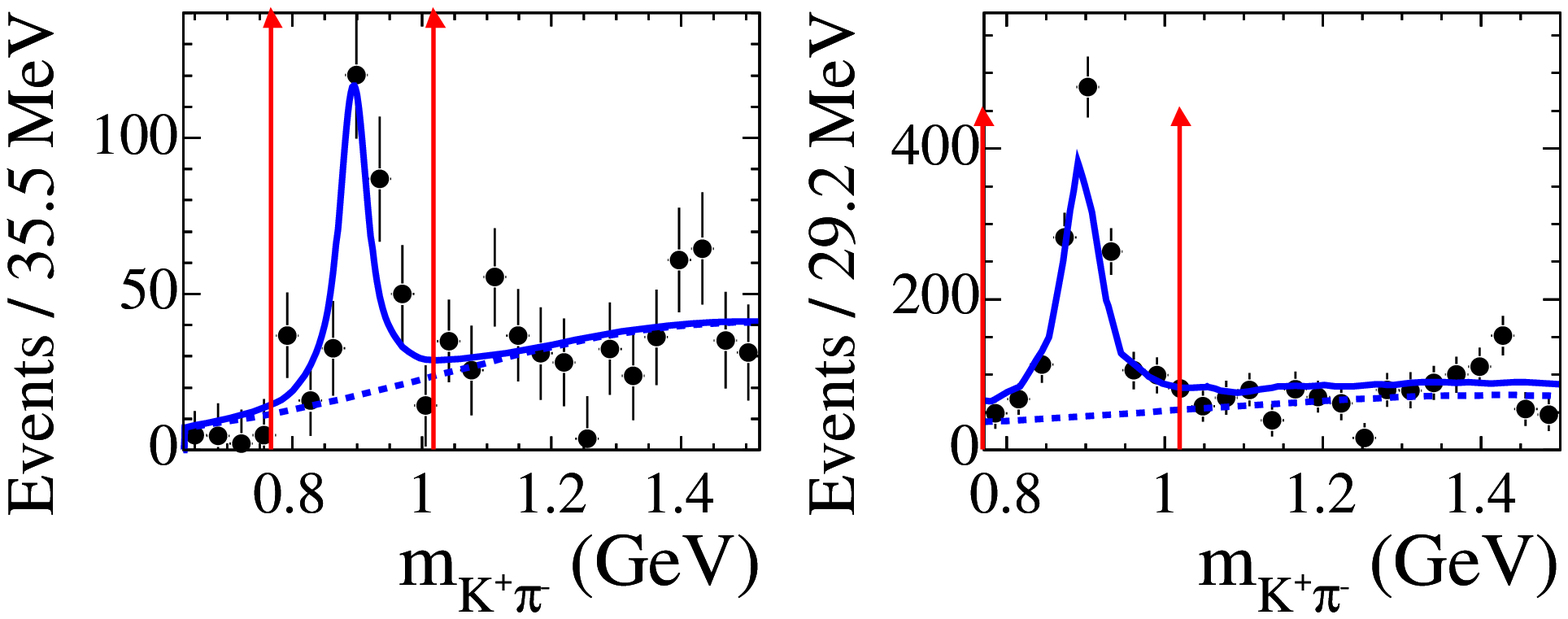}
\caption{\label{fig:mv_rhokst}
sPlots for the $\pi \pi$ (top) and $K \pi$ (bottom) invariant masses 
in the \brhopkz\ (left) and \brhozkz/\bfzkz\ (right) analyses. 
The points with error bars show the data.
The solid curve shows the signal and nonresonant background contribution,
the dashed curve is the nonresonant background contribution
(\rkp\ except for the top right plot where it represents the sum
of $f_0(1370) K^*$, $\pi \pi K^*$, and $\pi \pi K \pi$).
The arrows show the standard mass windows used in the final fit.}
\end{figure}

More recently \babar\ published 
an anlysis of all four \brhoks\ modes~\cite{bib:babarrhoks},
performed on a sample of 232 millions of \BB\ pairs.
It is based on an unbinned maximum-likelihood fit,
using seven variables: the $B$ meson energy-substituted mass \mes\ 
and energy difference \DE, 
a neural network output or a Fischer discriminant 
combining several event shape variables,
the two vector meson masses, and the two helicity angle cosines.
The fit allows the simultaneous extraction of the branching ratio 
and the fraction of longitudinal polarization.

\begin{table*}
\caption{\label{tab:rhoks}
Results from \babar\ on the \brhoks\ modes: 
signal yield with its statistical uncertainty,
significance (systematic uncertainties included), branching fraction
(90\% confidence level upper limit in parentheses),
fraction of longitudinal polarization and direct CP asymmetry.
(The numbers in brackets are not quoted as measurements.)}
\begin{ruledtabular}
\begin{tabular}{lrcccr}
Mode & Signal yield & Significance ($\sigma$) & {\cal B}($\times 10^{-6}$) & 
\fL\ & \Acp\ \\
\hline
\brhozkp  & $ 51\pm24$ & 2.5 & $<6.1 \, (3.6\pm1.7\pm0.8)$ & $[0.9\pm0.2]$ & \\
\brhomkp  & $ 60\pm24$ & 1.6 & $<12.0 \, (5.4\pm3.6\pm1.6)$ & & \\
\brhopkz  & $194\pm29$ & 7.1 & $9.6\pm1.7\pm1.5$ & $0.52\pm0.10\pm0.04$ & 
$-0.01\pm0.16\pm0.02$ \\
\brhozkz  & $185\pm30$ & 5.3 & $5.6\pm0.9\pm1.3$ & $0.57\pm0.09\pm0.08$ & 
$0.09\pm0.19\pm0.02$ \\
\end{tabular}
\end{ruledtabular}
\end{table*}

\begin{figure}[ht]
\includegraphics[width=0.4\textwidth]{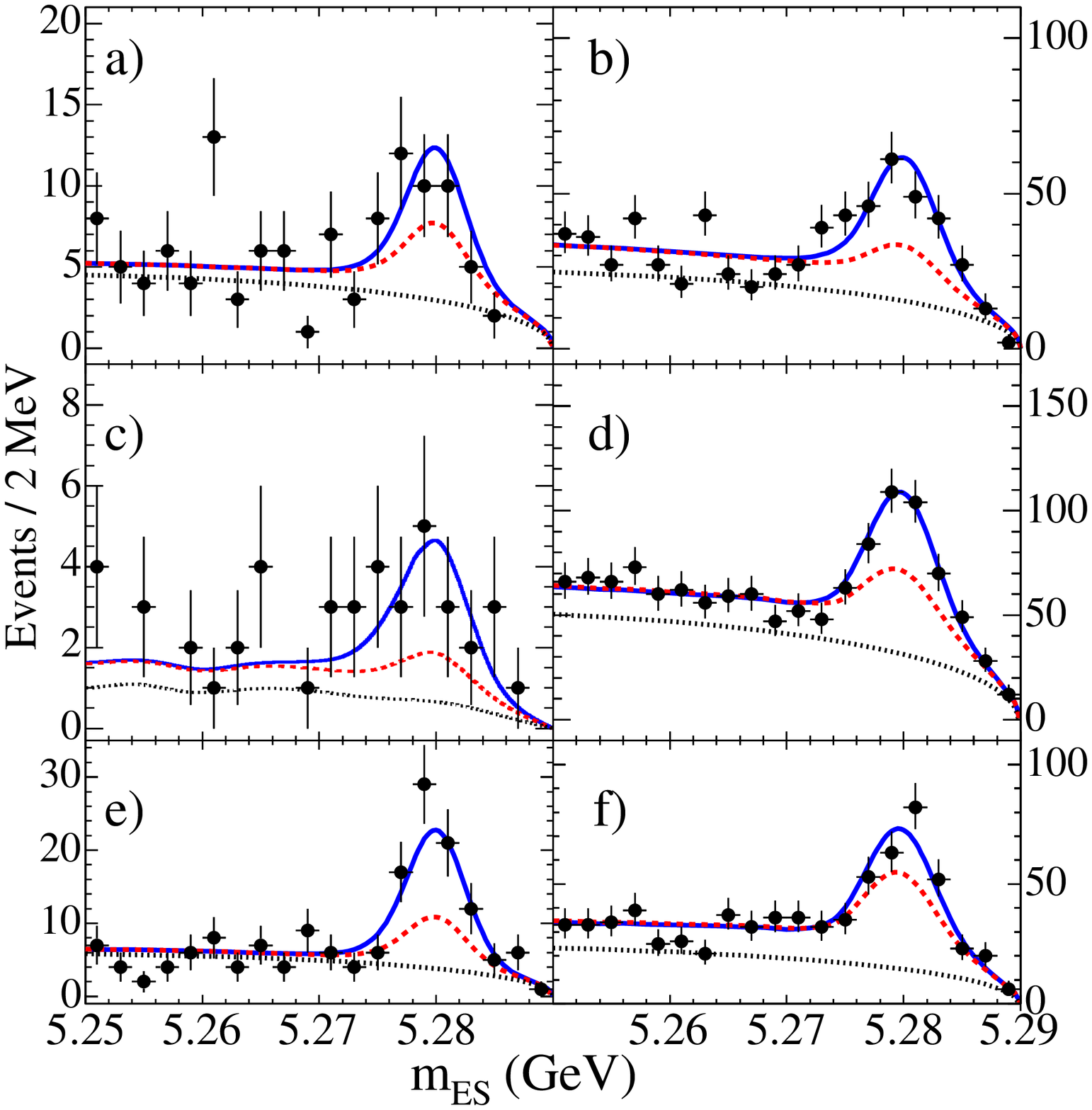}
\caption{\label{fig:mes_rhokst}Projections of \mes\ 
of events passing a signal likelihood threshold for
(a) \brhozkp, (b) \brhopkz, (c) \brhomkp, (d) \brhozkz, 
(e) \bfzkp, and (f) \bfzkz. 
The points with error bars show the data.
The solid curve is the fit function, 
the dashed curve is the total background contribution,
and the dotted curve is the continuum background contribution.}
\end{figure}

The major challenge in the analysis comes from the nonresonant backgrounds, 
which share the same final state as the signal.
They are studied by enlarging the vector meson mass windows,
as illustrated in Fig.~\ref{fig:mv_rhokst}.
As in Belle, a large \rkp\ background is seen in the \mkp\ distribution
in the \brhopkz\ mode.
The $K \pi$ system in this background is measured to be mostly S-wave.
The situation is even more complex in the \brhozkz\ mode,
since in addition to the \rkp\ background there are several contributions
seen in the \mpp\ distribution for a $\rho^0$ 
in contrast to the one for a $\rho^+$.
The $f_0(980)$ can be seen clearly.
In fact \bfzks, which is a scalar-vector mode, is considered 
as another signal to be measured in the same maximum-likelihood fit.
Also present are contributions from the $f_0(1370)$ 
and nonresonant $\pi \pi$.
The yields of the nonresonant backgrounds are fitted 
in the enlarged mass windows, then extrapolated to the standard ones
and fixed in the final fit with the standard mass windows.

The projection plots in the $B$ mass shown 
in Fig.~\ref{fig:mes_rhokst} illustrate the extraction of the signal
from the continuum and \BB\ backgrounds
in the four \brhoks\ channels and the two \bfzks\ modes.
Table~\ref{tab:rhoks} summarizes the results.
No significant enough signals are observed for \brhomkp\ and \brhozkp,
where upper limits at the 90 \% confidence level are set on the branching ratios.
For the latter a related signal \bfzkp\ is observed 
with a significance of 5.0~$\sigma$
and a measured branching fraction of 
$(5.2\pm1.2\pm0.5) \, 10^{-6}$.
In \brhopkz, the result is in very good agreement with the result from Belle,
with a similar precision.
The \brhozkz\ mode is observed for the first time.
The ratio between the branching fractions in these two modes is compatible
with the factor 2 expected from isospin symmetry.

The value of \Acp\ is measured in the two significant modes 
to be compatible with 0,
as expected since there is one dominant diagram.
Finally \fL\ is found close to 0.5 in these two modes. 
It is compatible with the measurement from Belle 
and has about the same precision.
It is again similar to the value found for $\phi K^*$.

\section{Modes with $\omega$}
\label{sec:om}

\begin{figure}[ht]
\includegraphics[width=0.4\textwidth]{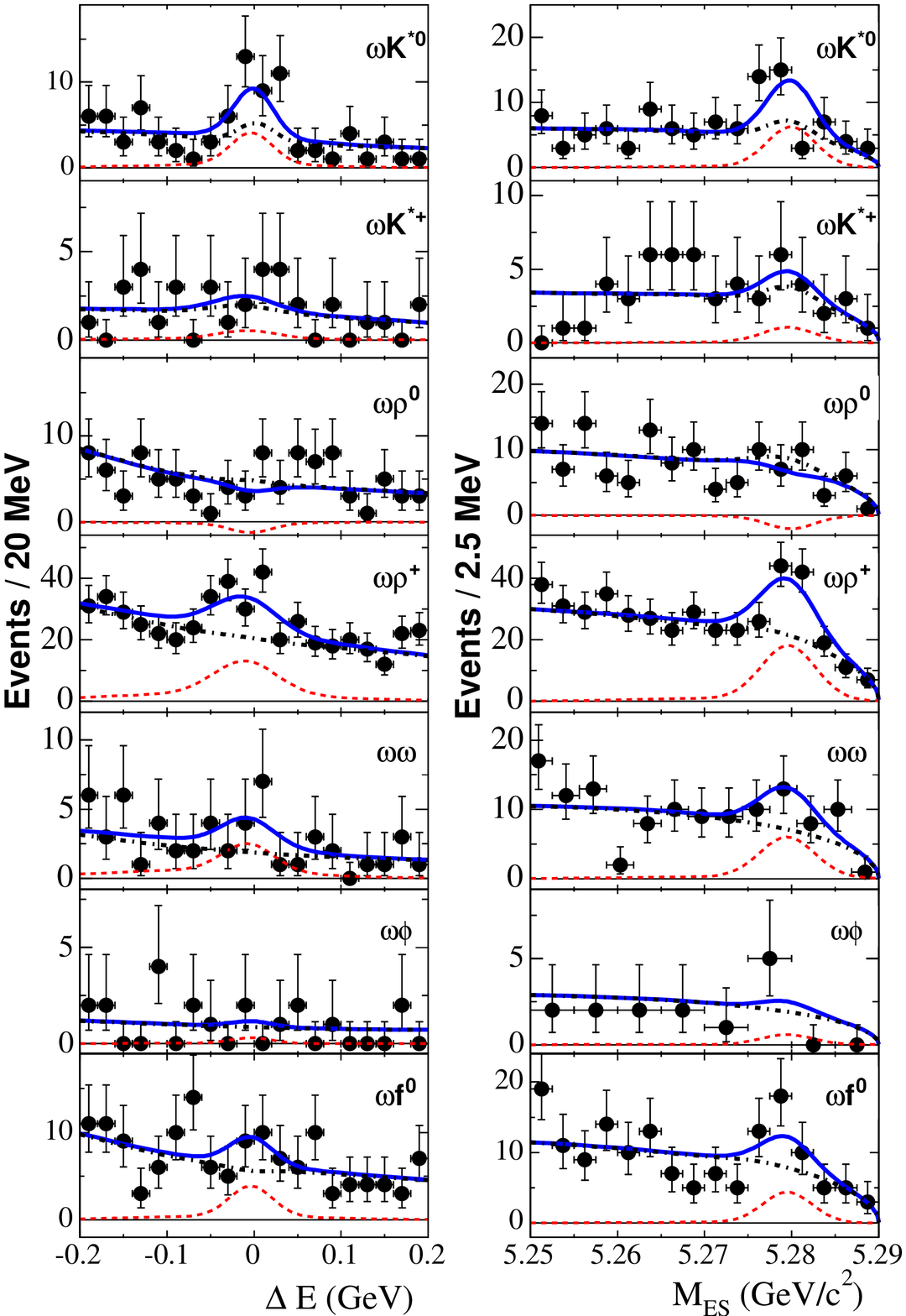}
\caption{\label{fig:demes_omega}Projections of \DE\ (left) and \mes\ (right) 
of events passing a signal likelihood threshold for, from top to bottom,
\bomkz, \bomkp, \bomrz, \bomrp, \bomom, \bomph, and \bomfz. 
The points with error bars show the data.
The solid curve is the fit function, 
the dashed curve is the signal contribution,
and the dot-dashed curve is the background contribution.}
\end{figure}

\begin{table*}[ht]
\caption{\label{tab:omega}
Results from \babar\ on modes involving an $\omega$ meson: 
signal yield with its statistical uncertainty,
significance (systematic uncertainties included), branching fraction
(90\% confidence level upper limit in parentheses),
fraction of longitudinal polarization and direct CP asymmetry.
(The numbers in brackets are not quoted as measurements.)
}
\begin{ruledtabular}
\begin{tabular}{lccccc}
Mode & Signal yield & Significance ($\sigma$) & {\cal B}($\times 10^{-6}$) & 
\fL\ & \Acp\ \\
\hline
\bomkz  & $ 55\pm20$ & 2.4 & $<4.2 \, (2.4\pm1.1\pm0.7)$ & $[0.71\pm0.25]$ & \\
\bomkp  & $  8\pm16$ & 0.4 & $<3.4 \, (0.6\pm1.3\pm1.0)$ & & \\
\bomrz  & $-18\pm16$ & 0.6 & $<1.5 \, (-0.6\pm0.7^{+0.8}_{-0.3})$ & & \\
\bomrp  & $156\pm32$ & 5.7 & $10.6\pm2.1^{+1.6}_{-1.0}$ & $0.82\pm0.11\pm0.02$ 
& $0.04\pm0.18\pm0.02$ \\
\bomom  & $ 48^{+24}_{-19}$ & 2.1 & $<4.0 \, (1.8^{+1.3}_{-0.9}\pm0.4)$ & 
$[0.71\pm0.25]$ & \\
\bomph  & $3.1\pm^{+4.4}_{-8.5}$ & 0.3 & $<1.2 \, (0.1\pm0.5\pm0.1)$ & & \\
\end{tabular}
\end{ruledtabular}
\end{table*}

\begin{figure}[ht]
\includegraphics[width=0.4\textwidth]{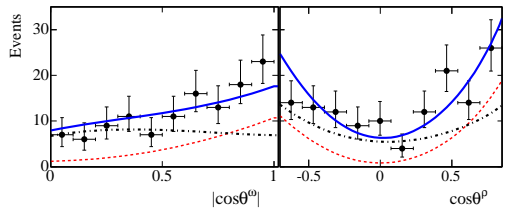}
\caption{\label{fig:hel_omega}Projections of the helicity-angle cosines 
for $\omega$ (left) and $\rho^+$ (right) of events passing 
a signal likelihood threshold from the fit for \bomrp\ decays.
The points with error bars show the data.
The solid curve is the fit function, 
the dashed curve is the signal contribution,
and the dot-dashed curve is the background contribution.}
\end{figure}

On the same sample of 232 millions of \BB\ pairs,
\babar\ has also recently published a search for several vector-vector modes
involving an $\omega$ meson~\cite{bib:babaromega}:
\bomkz, \bomkp, \bomrz, \bomrp, \bomom, and \bomph.
The related vector-scalar mode \bomfz\ was also searched for. 
An earlier search for \bomks\ and \bomrh\ on 89 millions of \BB\ pairs
resulted in the first observation of the \bomrp\ channel~\cite{bib:oldomega}.

The analysis is also based on an extended unbinned maximim-likelihood fit 
using the same seven variables as in the previous section.
Nonresonant $\pi \pi$ and $K \pi$ backgrounds are fixed in the fit
as determined from extrapolations from higher-mass regions.
The projection plots of \DE\ and \mes\ of Fig.~\ref{fig:demes_omega} illustrate
the extraction of the signal
from the continuum and \BB\ backgrounds in all these modes.
In most of them, no significant enough signal is seen.
The only channel where a significant signal is observed is \bomrp.
Its measured branching fraction is about 2 standard deviations smaller
than the one of $B^+ \to \rho^+ \rho^0$~\cite{bib:somov},
while these two branching fractions are naively expected to be equal.
Table~\ref{tab:omega} summarizes the results in all the modes.
To calculate the branching fraction, 
\fL\ is left free in the fit for the three modes
with a signal significance greater than 2$\sigma$ and is fixed otherwise.
Upper limits at the 90 \% confidence level are set on the branching fractions for the modes other than \bomrp.

The maximum-likelihood fit also provides the value of \fL\ in \bomrp,
which is found to be $0.82\pm0.11$, 
a high value expected for this tree-dominated mode.
This is illustrared in the projection plots of the helicity angle cosines
shown in Fig.~\ref{fig:hel_omega}.
The direct CP asymmetry is also measured and found to be compatible with 0. 

\section{Conclusion}
In summary, improved analyses 
with explicit consideration of nonresonant backgrounds have been performed
on several charmless hadronic vector-vector decays of the $B$ meson.
The \bomrp, \brhopkz, and \brhozkz\ modes have been observed
and measured in the past few years.
Improved upper limits have been set on the branching fraction 
of other vector-vector modes.

The recent results on vector-vector modes have also brought
more pieces to the polarization puzzle.
The penguin-dominated \brhopkz\ and \brhozkz\ modes have 
a fraction of longitudinal polarization of about 0.5 like $\phi K^*$,
while the tree-dominated \bomrp\ mode has one closer to 1 like $\rho \rho$.
As a lot of charmless vector-vector modes have not yet been observed,
new results can be expected with more data.

\end{document}